\documentclass[fleqn,usenatbib,a4paper]{mnras}

\usepackage[T1]{fontenc}
\usepackage{ae,aecompl}
\usepackage{color}

\usepackage{graphicx}	
\usepackage{amsmath}	
\usepackage{amssymb}	

\newcommand{\ltsima}{$\; \buildrel < \over \sim \;$}
\newcommand{\simlt}{\lower.5ex\hbox{\ltsima}}
\newcommand{\gtsima}{$\; \buildrel > \over \sim \;$}
\newcommand{\simgt}{\lower.5ex\hbox{\gtsima}}
\newcommand{\chandra}{{\it Chandra}}
\newcommand{\nustar}{{\it NuSTAR}}

\title[Seyfert galaxy NGC7130]{CO excitation in the Seyfert galaxy NGC7130}

\author[F. Pozzi et al.]{
F. Pozzi,$^{1,2}$\thanks{f.pozzi@unibo.it} L. Vallini,$^{3}$
C. Vignali,$^{1,2}$ M. Talia,$^{1}$ C. Gruppioni,$^{2}$
M. Mingozzi,$^{1,4}$\
\newauthor  M. Massardi,$^{5}$ P. Andreani$^{6}$
\\ 
$^{1}$ Dipartimento di Fisica e Astronomia, Universit\`a degli Studi di Bologna, 
Via Berti Pichat 6/2, I--40127 Bologna, Italy \\
$^{2}$ INAF --- Osservatorio Astronomico di Bologna, Via Ranzani 1,
I--40127, Italy
Bologna, Italy \\
$^{3}$ Nordita, KTH Royal Institute of Technology and Stockholm University,
Roslagstullsbacken 23, SE-106 91 Stockholm, Sweden \\
$^{4}$ INAF --- Osservatorio Astrofisico di Arcetri, largo E. Fermi 5,
I--50127, Firenze, Italy\\
$^{5}$ INAF-Osservatorio di Radioastronomia, Via Piero Gobetti 101, I-40129 Bologna, Italy\\
$^{6}$  European Southern Observatory, Karl-Schwarzschild-Strasse 2, D-85748, Garching, Germany\\
}

\date{Accepted XXX. Received YYY; in original form ZZZ}

\pubyear{2015}

\begin{document}
\label{firstpage}
\pagerange{\pageref{firstpage}--\pageref{lastpage}}
\maketitle

\begin{abstract}

We present a coherent multi-band modelling of the CO Spectral Energy Distribution of the local Seyfert Galaxy
NGC7130 to assess the impact of the AGN activity on the molecular gas.
We take advantage of all the available data from X-ray to
the sub-mm, including ALMA data. The high-resolution ($\sim{0.2}^{"}$) ALMA CO(6-5) 
 data constrain the spatial
extension of the CO emission down to $\sim$70 pc scale. From the
analysis of the archival \chandra\ and  \nustar\ data,
 we infer the presence of a buried, Compton-thick AGN of moderate
 luminosity, $L_{2-10keV}{\sim}1.6{\times}10^{43}$~erg~s$^{-1}$.
We explore photodissociation and X-ray-dominated regions (PDRs and XDRs) models to reproduce the CO
emission. We find that PDRs can reproduce the CO lines up to
J${\sim}$6, however, the higher rotational ladder requires the presence
of a separate source of excitation. We consider X-ray heating by the AGN as a source of excitation, and
find that it can reproduce the observed CO Spectral Energy
Distribution. By adopting a composite PDR+XDR model, we derive
molecular cloud properties.
Our study clearly indicates the capabilities offered by
current-generation of instruments to shed
light on the properties of nearby galaxies adopting state-of-the art
physical modelling.

\end{abstract}




\section{Introduction}
\label{intro}
In the far-IR and sub-mm range there are several molecular and atomic
lines that probe the different phases of the interstellar medium (ISM)
and whose study gives important constraints on its physical condition.
In particular, the carbon monoxide (CO) molecule is a good tracer of
the molecular gas phase being the second most abundant molecule after
the molecular hydrogen (H$_{2}$). 

The CO spectral line energy distribution (i.e. CO SLED) 
 is sensitive to the molecular gas kinetic
 temperature and can provide constraints on the dominant heating
 mechanism, controlled by radiative heating from stars and Active
 Galactic Nuclei (AGN) (e.g. \citealt{2009ApJ...702.1321O}). While the CO SLED line fluxes usually rise up to the CO(5-4)
transition and then decrease for typical region around newly formed
stars (`Photon Dissociated Region', PDR), the high rotational
transition levels, if excited, trace the presence of warm gas ($100~K$<$T$<$1000~K$) heated by X-ray
photons (`X-ray Dominated Regions', XDRs, e.g. \citealt{2010A&A...518L..42V}) or by shocks
(e.g. \citealt{2010A&A...518L..37P}, \citealt{2013ApJ...762L..16M}). This has been shown firstly with observations from ground--based 
telescopes, mainly for high-$z$ galaxies and AGN, taking advantage of the possibility
to probe the rest-frame mid/high-J transitions ($\nu$>500-600 GHz) with
sub-mm/mm facilities such as the IRAM and CARMA interferometers
(i.e. \citealt{2005ARA&A..43..677S}).
Subsequently the Herschel satellite
has extended the CO detections to high transitions
levels also in the local Universe, using both the PACS and SPIRE/FTS spectrometers
(\citealt{2013ApJ...768...55P} and \citealt{2015ApJ...801...72R}).
With the advent of the sub-mm/mm ALMA telescope, a further step has
been done in this field, thanks to its high instantaneous
 sensitivity and spatial resolution. In the local Universe, regions of the order of 50-100 pc can be easily
resolved supplying crucial information on the modelling.

NGC7130 is a local ($z$=0.0161) Luminous Infrared Galaxy ($L_{IR}{\sim}10^{11.3}$
$L_{\odot}$, see \citealt{2016MNRAS.458.4297G}) with an ``ambiguous'' 
nature. 
On the one hand, it was optically morphologically classified as a peculiar Sa spiral galaxy  (\citealt{1998ApJ...506..673L}) and a Seyfert 1.9 source from the 
optical emission line ratio (\citealt{2006A&A...455..773V}). 
Recently \cite{2016MNRAS.458.4297G} from a 
SED-decomposition analysis found that the data were well reproduced without the need of an AGN torus component. 
On the other hand, the presence of a buried or low-luminosity AGN is suggested by the 
fine structure IR lines and X-ray data. In the mid-IR, \cite{2010ApJ...709.1257T} 
detected [NeV] lines at 14.32${\mu}$m and 24.31${\mu}$m which, due to
their high ionization potential of 97 eV, can be reasonably excited
only by an AGN.  In the far-IR, \cite {2015ApJ...799...21S} found a low [OIII]88${\mu}$m/[OIV]26${\mu}$m ratio
(${\sim}$1), similar to other Seyfert rather than to starbust
galaxies. In the X-ray band, \cite{2005ApJ...618..167L}, analyzing \chandra\ data, 
claimed the presence of a buried Compton-thick AGN ($N_{H}$>$10^{24}$~cm$^{-2}$) with observed low luminosity 
($L_{0.5-10}{\sim}10^{41}$ erg~s$^{-1}$), dominating at energies >2 keV. 
From the K$\alpha$ iron emission line, the same authors estimated 
an intrinsic 2--10~keV luminosity of ${\sim}10^{43}$~erg~s$^{-1}$. Indications 
of extreme obscuration towards this source came also from \cite{1999ApJ...522..157R} 
on the basis of the 2--10~keV/[OIII] flux ratio. 
Summarizing, NGC7130 seems a prototype of a strong starburst
probably hosting a Compton-thick AGN.

The goal of this letter is to definitively assess the impact of the
AGN activity on the molecular gas in NGC7130. This will be achieved
via the modelling of the NGC7130 CO SLED. A coherent multi-phase (and
multi-band) analysis of NGC7130 build upon literature data and new
X-ray observations will be presented.
The paper is structured as follows: in Sec. \ref{data} we provide
details on the new \chandra\ and \nustar\ data, and we highlight the importance of the
high-resolution ALMA archival observations in devising a precise CO
SLED model. In Sec. \ref{modeling} we present the model assumptions
and the results. Finally, in Sec. \ref{summary} we draw a short summary.

\section{{\bf Data}}
\label{data}

\begin{figure}
\includegraphics[width=0.45\textwidth,height=6cm,angle=0]{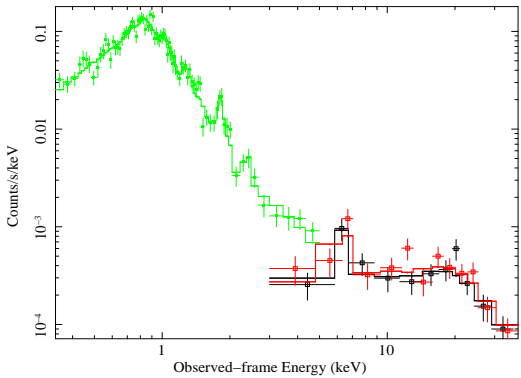}
\caption{\chandra\ (green points) and \nustar\ FPMA and FPMB (black and red points, respectively)  
spectra of NGC7130 and best-fit modelling (see text for details). }
\label{figure_nustar}
\end{figure}

\subsection{Chandra and NuSTAR data}
\label{xray_section}
NGC7130 was observed by \chandra\ in Cycle~2 with the ACIS-S3 CCD at the aimpoint (on Oct. 23$^{rd}$, 2001), 
for an exposure time of 38.6~ks. The source shows a point-like emission surrounded by diffuse emission, as estensively described 
by \cite{2005ApJ...618..167L}. We extracted the \chandra\ spectrum using a circular region of 15\arcsec\ radius (to include the nuclear and 
the diffuse emission and have enough photon statistics for a proper spectral modelling), and the background from 
a circular region (radius of 22\arcsec) with no sign of diffuse emission. The spectrum, of about $\approx$3600 net counts, was rebinned to have at least 25 counts per bin to apply $\chi^{2}$ statistics within the {\sc xspec} package \citep{1996ASPC..101...17A}.

\nustar\ observed NGC7130 on Aug. 17$^{th}$, 2014, for an exposure time of 21.2~ks. Data were reprocessed and screened using 
standard settings and the {\sc nupipeline} task, and source and background spectra (plus the corresponding response matrices) were 
extracted using {\sc nuproducts}. Given the relatively low flux of the source, a circular extraction region of 30\arcsec\ radius 
was selected, while background was extracted from two nearby circular regions of 40\arcsec\ radius. The source net counts, $\approx$190 and $\approx$220 in the FPMA and FPMB cameras, respectively, were rebinned to 20 counts per bin. The source signal is detected up to $\approx$30~keV.

\chandra\ and \nustar\ data were fitted simultaneously with the same modelling, which accounts for the dominant thermal emission in 
the soft X-rays (\chandra\ band) and for the nuclear emission at hard X-ray energies (\nustar\ range). In particular, data below a few keV 
are modeled with two thermal components ({\sc mekal} model within {\sc xspec}), similarly to \cite{2005ApJ...618..167L}. 
The derived plasma temperatures are $kT_{1}=0.59^{+0.04}_{-0.05}$~keV and $kT_{2}=1.58^{+0.93}_{-0.37}$~keV. 
Four additional lines, likely related to iron and SiXIII transitions, seem to be required (at a statistical level of at least 90\%).
%
%
The \chandra\ data appear very noisy above $\approx$5~keV and were therefore removed; however, in this energy range \nustar\ data, albeit characterized by a lower spectral resolution ($\approx$400~eV vs. 130~eV typically provided by \chandra\  at 6~keV), are fundamental to characterize the iron K$\alpha$ emission line, define the intrinsic continuum of NGC7130 and place constraints on the level of obscuration. 
To achieve these goals, we used the {\sc MYTorus} model \citep{2009MNRAS.397.1549M}, which is based on Monte Carlo simulations and, 
assuming a toroidal geometry for the reprocessor (uniform and cold), self-consistently includes reflection and scattering. 
While the photon index was fixed to 1.9, as typically observed in AGN
and quasars, the column density 
of the torus and the inclination angle between the observer's line-of-sight and the symmetry axis of the torus $\theta_{obs}$ are left as free parameters of the fit. We obtain an equatorial column density of
($5.3\pm{1.6}){\times}10^{24}$~cm$^{-2}$ and
$\theta_{obs}=79\pm{5}$~degrees. This overall modelling provides a
$\chi^{2}$/dof (degrees of freedom) of 99.1/85; most of the deviations
from the best-fit model appear in the soft band and are of limited
relevance for the purposes of the present X-ray analysis.
The intrinsic 2--10~keV and 1--100~keV luminosities of the AGN are $1.6{\times}10^{43}$~erg~s$^{-1}$ and $5.1{\times}10^{43}$~erg~s$^{-1}$, respectively. 

Our results are in good agreement with the analysis of the hard-X
  ray data (\nustar\ and {\it Swift/BAT}) 
from \cite{2016ApJ...825...85K}, where NGC7130 was defined as a
moderate CT-AGN, and with the recent work of \cite{2017MNRAS.tmp..179R},
that, performing an analysis similar to ours, confirms the presence
of CT obscuration.
 
\subsection{CO data}
\label{alma_sec}

\begin{figure}
	\includegraphics[width=7.6cm,height=5.cm]{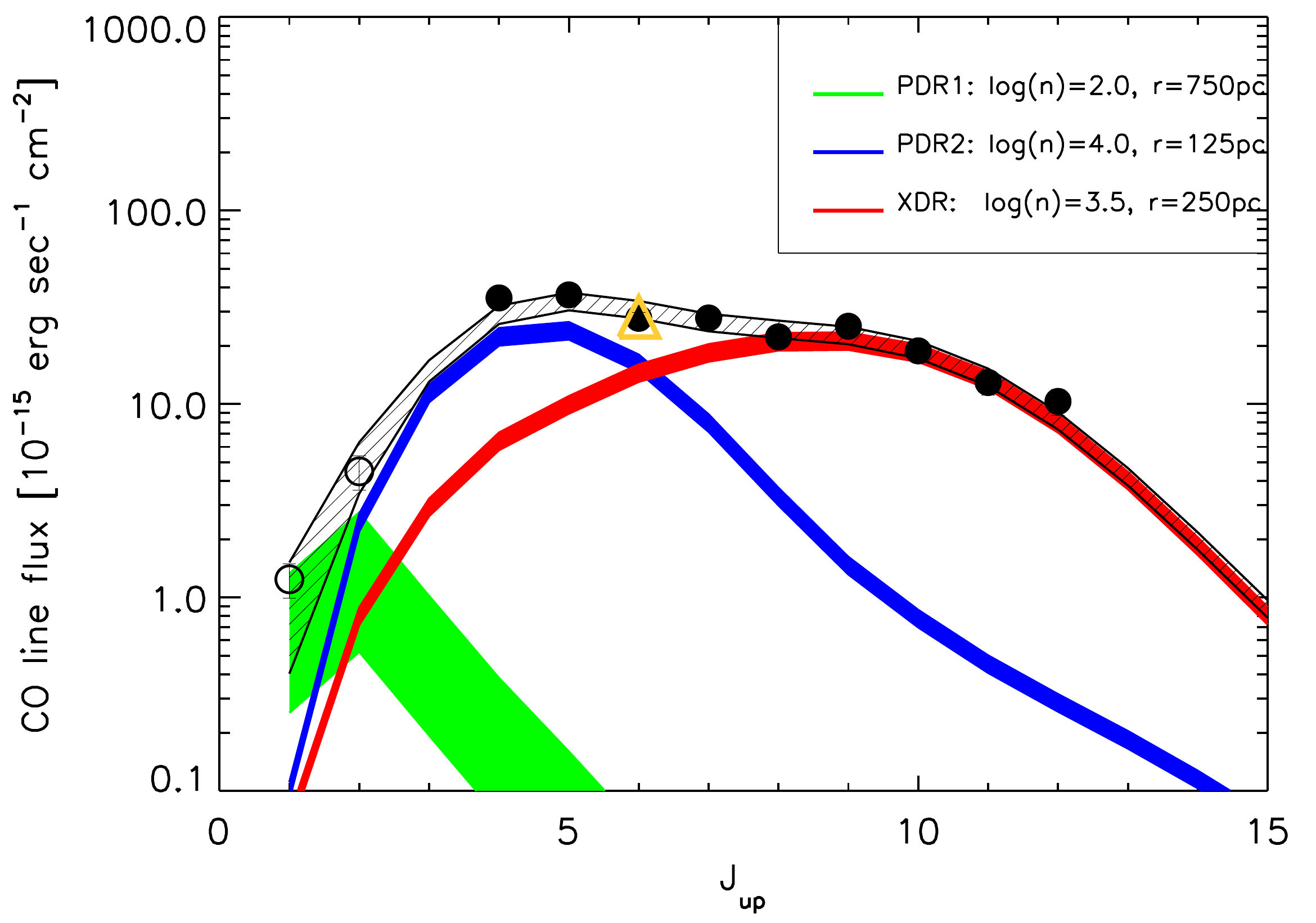}
    \caption{Observed CO SLED and best-fit model
      over-plotted. The low-J transition (open circles) are from the
      SEST telescope, while the mid/high-J
 transition levels (filled circles) are from Herschel SPIRE/FTS observations. The
 CO(6-5) ALMA data is highlighted as a yellow triangle. The green, blue
 and red lines correspond to the three model components (parameters reported
 in the legend). The black line is the sum of the three components. The thickness of the colored
      regions represents the 1$\sigma$ uncertainties. }
    \label{figure_co}
\end{figure}

The observed CO SLED is shown in Fig. \ref{figure_co}.
 The mid/high-J levels ($J_{up}$>4) are from Herschel SPIRE/FTS observations
 (\citealt{2013ApJ...768...55P}), while the low-J transitions are from \cite{2007A&A...462..575A} using the
 Swedish ESO Submillimeter Telescope (SEST).
 The Herschel observations are performed to cover a field of view
 (FOV) of ${\sim}2^{\prime}$. The observations performed with the single dish SEST
 Telescope are pointed observations with a FOV (${\sim}2{\times}$FWHM)
 of ${\sim}1^{\prime}$. NGC7130 appears point-like
 (\citealt{2013ApJ...768...55P}) in the Herschel SPIRE 
 photometric bands (from 250 to 500 ${\mu}$m) characterized by a maximun
 FWHM of 40$^{\prime\prime}$. Assuming that the dust (sampled by the SPIRE photometric observations) and the gas (sampled by SPIRE/FTS
 observations) are almost co-spatial, this means that all the CO fluxes
 correspond to the integrated emission of the galaxy.

The yellow triangle represent the new ALMA CO(6-5) data from \cite{2016ApJ...820..118Z}.
For details on the analysis of the ALMA data we
refer to \cite{2016ApJ...820..118Z}; here we only remind the points
which are relevant for our analysis. The ALMA observations have been performed
in Cycle 2 (project  2013.1.00524.S, PI: Nanyao Lu) with the Band 9
receiver searching for the CO(6-5) transition at 691.473 GHz. The
configuration user (C34-5) allowed a synthesized beam FWHM of
${\sim}0.20{\times}0.14^{\prime\prime}$ corresponding to physical
scales of ${\sim}70{\times}49$ pc.
 The total line flux is 1230${\pm}$74 Jy km
s$^{-1}$ and half of this flux in contained in a nuclear region of diameter
${\sim}500$ pc, while the rest in structures at distances of ${\sim}500$
pc from the center. The ALMA flux is completly consistent with
  the Herschel measurement (see also \citealt{2017arXiv170300005L}).  
This suggests that the ALMA observation has collected all the CO(6-5)
flux, despite the largest angular scales recovered by the ALMA configuration is only
${\sim}3^{\prime\prime}$, corresponding to ${\sim}$1 kpc.

\section{ISM modelling}
\label{modeling}

We model the CO SLED using the version c13.03 of the
photoionization/photodissociation code CLOUDY
(\citealt{2013RMxAA..49..137F}) in order to infer the conditions of
the molecular gas where the different CO transition lines are excited.
In this work, we run two sets of CLOUDY models that
  assume the molecular clouds to be a 1-D gas slab with constant
  density located at a fixed distance $d$ from the source of the
  illuminating radiation (e.g. \citealt{2017MNRAS.tmp..185V}). We adopt different prescriptions for the
  spectral energy distribution (SED) of the impinging radiation
  field. The first set of runs (PDR models, hereafter) assumes a pure
  stellar SED illuminating the gas slab. The SED is obtained with the
  stellar population synthesis code Starburst99
  (\citealt{1999ApJS..123....3L}), assuming a Salpeter (\citealt{1955VA......1..283S}) 
Initial Mass Function (IMF) in the
  range 1-100 $M_{\odot}$, and Lejeune-Schmutz stellar atmospheres
  (\citealt{1992PASP..104.1164S}; \citealt{1997A&AS..125..229L}). We
  adopt $Z_* = Z_{\odot}$, and a continuous star formation normalized
  to reproduce the NGC7130 of SFR${\sim}$21 M$_{\odot}$yr$^{-1}$
  (\citealt{2016MNRAS.458.4297G}). The second set of runs (XDR
  models, hereafter) follows the same approach proposed by \cite{2008ApJ...678..686A} for modeling the AGN radiation impinging a gas slab. More
  precisely, to account for the radiation field from the AGN, we
  adopt the \texttt{table AGN} command in CLOUDY
  (\citealt{1997ApJ...487..555K}). 
This command simulates an UV bump ($f \propto \nu ^{-0.5}
exp(-h\nu/k T_{cut})$) plus an X-ray power-law ($\nu^{-1}$). For the
value of $T_{cut}$, we use the value $10^6\, \rm K$ proposed by 
\cite{1997ApJ...487..555K} and adopted in \cite{2008ApJ...678..686A}. The ratio of the UV and X-ray continua ($a_{ox}$) is
  fixed to $-1.4$, a typical value for an AGN
(\citealt{1981ApJ...245..357Z}). We normalize the spectrum to match
the observed  $L_{1-100keV}{\sim}5{\times}10^{43}$ erg s$^{-1}$ (see Sec.\ref{xray_section}).

For the PDR models, we run a total of 35 CLOUDY
  simulations leaving (i) the gas density $n$, and (ii) the distance
  $d$ between the radiation source and the cloud free to vary in the
  range $log(n/\rm cm^{-3})=[2- 4.5]$ and $d=[75-250]$ pc,
  respectively. 
Note that a variation in $d$ translates into different values for the
impinging far-ultraviolet flux (FUV $6-13.6 \,\rm{eV}$, for the PDR
models), and X-ray 1-100 keV flux ($f_x$, for XDR models) at the cloud
surface. The incident FUV flux (X-ray flux) is a key parameter
influencing the heating and cooling processes in the photodissociation
(X-ray dominated) regions (e.g. \citealt{1999RvMP...71..173H},
\citealt{2007A&A...461..793M}). The former is usually normalized to
the solar neighborhood value ($1.6{\times}10^{-3} \rm{erg\,
  s^{-1}\, cm^{-2}}$, \citealt{1968BAN....19..421H}) and is indicated with
$G_0$.  The chosen density range allows to sample
  the typical densities observed in giant molecular clouds (GMCs,
  e.g. \cite{2007ARA&A..45..565M}, and references therein). 
To constrain the parameter $d$, we take advantage of the
key information provided by the ALMA observations. As a matter of
fact, $d$ can range from the minimum spatial scale recovered by ALMA
(${\sim}$70 pc, see \ref{alma_sec})  to the maximum extension of the ALMA CO(6-5) detection (radius $\sim$250 pc; see
  Sec. \ref{alma_sec}).  We simulate also two distances larger than
  the CO(6-5) emission ($d$=[500,750] pc). Indeed, the constraint relative to the
  extension of the CO emission (concerning the CO(6-5),
  and likely also the higher and denser J-transitions) is not applicable to the lower and more diffuse transitions, given the
  spatial resolution of the Herschel and SEST observations
  (see Sec. \ref{alma_sec}). For the XDR models we keep the same range for $d$ adopted for the PDR
 models (up to $d$=$250$ pc), while we consider a slightly higher density range, i.e. $log(n/\rm cm^{-3})=[3.5 - 5.5]$.
With our choice of $d$, we cover the range
$G_0=(4.9-410){\times}10^{2}$ and $f_x=0.3-80$ in $\rm{erg\, s^{-1}\,
  cm^{-2}}$.

 For both the PDR and XDR models, CLOUDY computes the  emergent CO
 line emissivity ($\epsilon_{CO,J-J-1}$, with $J$ varying from 1 to
 11) as a function of the depth the radiation penetrates into the cloud. In our models we set a maximum total gas column density $log(N_H/{\rm cm^{-2}})$=23.


{In Fig. \ref{figure_contours} we show the CO(5-4)/CO(4-3) (top panel) and the CO(10-9)/CO(5-4) (botton panel) ratios as a function of $n$ and $N_{H}$, obtained with a set of PDR models fixed $d=250$ pc, from the stellar radiation source. This spatial scale is equal to the extension of the CO(6-5) emission observed with ALMA by \cite{2016ApJ...820..118Z}.
The simultaneous study of the two ratios is essential to understand whether the CO SLED can be reproduced by a unique set of PDR models with clouds at fixed distance $d$ from the galaxy center, or additional components are needed.
In  Fig. \ref{figure_contours} (top panel), the white line highlights the parameter space in $n$ and $N_{H}$,
that can reproduce the observed CO(5-4)/CO(4-3)) ratio. These parameter ranges, typical of PDRs,
do not reproduce the observed CO(10-9)/CO(5-4) ratio. Hence, for reproducing the mid to high-J CO transition
there are two possibilities: (a) a PDR + a component that accounts for the contribution of
the shock--heated gas and (b) a PDR + an XDR.  We favor the
scenario (a) (PDR+XDR), as the most likely one, on the basis of the following
considerations. 


\begin{figure}
	\includegraphics[width=9cm,height=8cm]{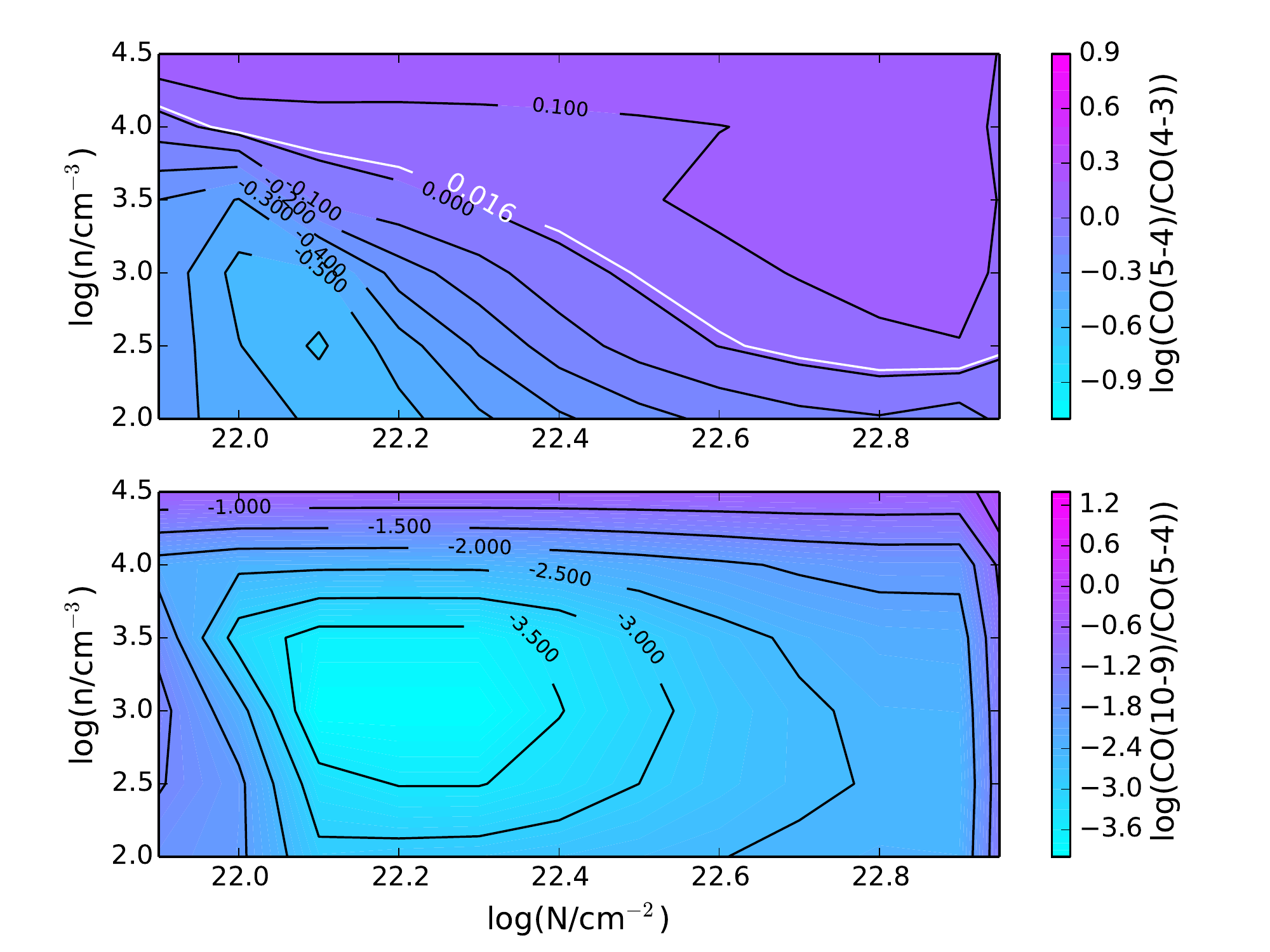}
    \caption{CO(5-4)/CO(4-3) ({\it Top panel}) and
      CO(10-9)/CO(5-4) ({\it Botton panel}) ratios as a function of
      the density $n$ and column density $N_H$ for PDRs. The distance
      is fixed ($d$=250 pc). The white line ({\it Top panel}) highlights the parameter space corresponding to the
      observed ratio (log(CO(5-4)/C
O(4-3))=0.017). The observed
      log(CO(10-9)/CO(5-4)) ratio of -0.29 is not reproduced ({\it Botton panel}).}
    \label{figure_contours}
\end{figure}

A chemistry driven by shocks (scenario (b), supported by \citealt{2014A&A...566A..49P}) is 
unlikely given the observed $L_{CO}/L_{IR}$ ratio. As reported in
\cite{2014MNRAS.445.2848G}, a key difference between shocks and radiative excitation
by UV and/or X-rays is that shocks do not heat dust as effectively as they do for gas. NGC7130 has an observed 
$L_{CO}/L_{FIR}$ ratio ${\sim}1.5{\times}10^{-4}$, a factor typical of
XDR/PDR models (e.g. \citealt{2005A&A...436..397M}), while a value of ${\sim}7{\times}10^{-4}$, as found
by \cite{2013ApJ...762L..16M} in NGC6240, seems to be more related to shocks, 
compressing the gas and heating it to higher temperatures, while not
affecting the dust (see also \citealt{2016A&A...596A..85L}). Moreover, X-ray photons, as the primary heating
mechanism of the high-J CO lines is supported by (i) the clear
detection, thanks mostly to \nustar\ data, of
a hard X-ray flux originating from a central accreting engine (see
Sec. \ref{xray_section}) and (ii) by the presence of mid-IR [NeV] that
is a clear sign (\citealt{2010ApJ...709.1257T}, \citealt{2016MNRAS.458.4297G})
  of the effect of AGN activity on the ISM of the host galaxy. 




 Given the considerations reported above, we consider three components: a PDR and one XDR for the
mid to high-J CO transitions, embedded in a more rarefied and extended PDR
envelope accounting for the low-J transition. A similar model
was presented by \cite{2010A&A...518L..42V} for reproducing the
CO SLED of Mrk231. In  Fig. \ref{figure_co} our best-fit model is
presented. We find the best fit by a ${\chi}^{2}$ minimitation routine with 9
free parameters: the density, the radiation field intensity and the
normalization for each of the 3 components. We fix the column
density $N_{H}~{\sim}~10^{22}$cm$^{-2}$ for the PDRs and
$N_{H}~{\sim}~10^{23}$cm$^{-2}$  for XDR.
As shown by \cite{2005A&A...436..397M} (see also
  \citealt{2007ARA&A..45..565M}, \citealt{2014A&A...564A.121C}), while in PDR
the carbon monoxide over atomic carbon ratio (CO/C) ratio raises by 4 dex from
the illuminated cloud surface to $N_{H}{\sim}10^{22}$cm$^{-2}$, in XDR CO/C
is almost constant up to $N_{H}{\sim}10^{22}$cm$^{-2}$ and then
increases slowly.

sThe $1{\sigma}$ level of confidence on the normalizations reported in
Fig. \ref{figure_co} have been obtained by marginalizing over the
other parameters and considering a ${\Delta}{\chi}^{2}$=3.5 (see
\citealt{1976ApJ...208..177L}). According to our model, the PDR and XDR reproducing the mid-J and high-J
transitions are characterized by gas density $log(n/cm^{-3})=4.0$
(3.5) and the gas is located at 125 pc (250 pc) from the source of
radiation, respectively (corresponding to $G_{0}=4.4{\times}10^{3}$ and
$f_x$=5.3 erg cm$^{-2}$s$^{-1}$). The typical thickness and gas
temperature of the CO
emitting regions for the PDR and XDR are: $l{\sim}$0.3, 4 pc and T=40,
T=100 K, respectively. As expected X-rays
photons keep the gas temperature higher at larger depths. 
The diffuse PDR component, accounting for the low-J lines, is
characterised by a low-density gas ($log(n/cm^{-3})=2$)
 exposed to a FUV flux $G_{0}=4.9{\times}10^{2}$. The emission
 originates from regions of thickness $l$=20 pc, typical of GMC in the
 MW (see \citealt{2011ApJ...729..133M}).
The gas temperature is T=25 K. 

The presence of, at least, two temperatures components to model the 
  CO SLED of NGC7130 is consistent with the results found in star-forming galaxies 
 (i.e. \citealt{2015A&A...577A..46D},
 \citealt{2016ApJ...829...93K}). The presence of an extra XDR component extends to
 lower X-ray luminosities (by a factor of ${\sim}$4) the result found by \cite{2010A&A...518L..42V} for
 modeling the high-J levels of the local CT-AGN Mrk231.

The total cold gas mass necessary to fit the CO observation is
$M_{cold}~{\sim}~10^{9}M_{\odot}$ and completely dominating the total mass.
Considering the observed CO(1-0) luminosity
($L_{CO}=10^{8.8}$ pc$^{2}$~K~km~s$^{-1}$), the estimated gas mass
allow us to estimate directly the CO-to-H$_{2}$ conversion factor, $\alpha_{CO}$=1.1 M$_\odot$
pc$^{-2}$~K~km~s$^{-1}$ (where
$M_{cold}=\alpha_{CO}L_{CO}$, see
\citealt{2005ARA&A..43..677S}). The CO-to-H$_{2}$ conversion
factor shows a wide range of values reflecting the different ISM conditions (see Fig. 12 from the review of
\citealt{2013ARA&A..51..207B}). The normal galaxy
disks return high values for $\alpha_{CO}$ ($\alpha_{CO}$=4.6 M$_\odot$
pc$^{-2}$~K~km~s$^{-1}$ for the MW,
\citealt{1991IAUS..146..235S}) while $\alpha_{CO}$ drops to very low values (up to
$\alpha_{CO}$=0.3 M$_\odot$ pc$^{-2}$~K~km~s$^{-1}$) for active and ULIRGs galaxies
(see also \citealt{1998ApJ...507..615D},
\citealt{2012MNRAS.426.2601P}). The value found is
consistent with the overall properties of NGC7130 being a LIRG galaxy
hosting a moderate, obscured AGN.
 
\section{Summary}
\label{summary}

In this letter, we have presented a detailed modelling of the molecular gas in NGC7130 through the study of the CO Spectral Line Energy Distribution.
Key informations are derived from the ALMA data of the CO(6-5) transition and from the X-ray \chandra + \nustar\ data. The
high-resolution ALMA data allow us to
constrain the physical CO emission down to ${\sim}$70 pc scales, while
the X-ray data confirm the presence of a central highly obscured AGN.


According to our modelling, three components are needed to reproduce
the CO SLED: a PDR with parameters typical of the diffuse cold ISM
($log(n/cm^{-3})=2.0 $, T${\sim}$25 K); a PDR, at
relatively higher density and temperatures ($log(n/cm^{-3})=4.0 $,
T${\sim}$40 K) and an extra warm XDR component for ($log(n/cm^{-3})=3.5 $, T${\sim}$100 K). 
An XDR driven chemistry for the high-J CO lines is supported by our
multicomponent  $\chi^2$--fitting the CO SLED. 
The presence of an XDR is discussed against heating due to shocks.
Our modeling predicts a cold gas mass of $M_{cold}~{\sim}~10^{9}M_{\odot}$,
that implies a CO-to-H$_{2}$ conversion factor $\alpha_{CO}$=1.1 M$_\odot$ pc$^{-2}$~K~km~s$^{-1}$.

These results show the strength of combining multi-band and multi-resolution data to assess the impact of the
AGN and star-formation activity on the physics of molecular gas.
The future exploitation of the data in the ALMA archive will allow
 us to enlarge the sample and place the results on NGC7130 on a more statistically significant
context. Moreover, we intend to extend the analysis performed on the CO to other
molecules, such as HCN, HCO$^{+}$. These molecules, characterized by high critical
densities ($log(n/cm^{-3})$\gtsima 5), allow to trace different parameter
space of the GMCs, and their flux ratios permit to highlight the presence and strength of an AGN (i.e. \citealt{2016AJ....152..218I}).
\section{Acknowledgements}
We acknowledge support from Italian node of the Alma Regional Center.
This paper makes use of the ALMA data ADS/JAO.ALMA\#2013.1.00524.S. FP
kindly thanks the referee Matthew Malkan and Andrea Cimatti.





\bibliographystyle{mn2e}
\bibliography{pozzi}







\bsp	
\label{lastpage}
\end{document}